\documentclass[twocolumn,superscriptaddress,floatfix]{revtex4}

\usepackage{graphicx}
\usepackage{dcolumn}
\usepackage{bm}
\usepackage{xcolor}

\begin{document}


\title{Spin channel induced directional dependent spin exchange interactions between divacantly substituted Fe atoms in graphene}

\author{R. Putnam}
\affiliation{Department of Physics, University of North Florida, Jacksonville, FL 32224, USA}
\author{A. V. Balatsky}
\affiliation
{Nordic Institute for Theoretical Physics, KTH Royal Institute of Technology and Stockholm University, Roslagstullsbacken 23, 106 91 Stockholm, Sweden}
\affiliation
{Department of Physics, University of Connecticut,
Storrs, CT 06269, USA}
\author{J. T. Haraldsen}
\affiliation{Department of Physics, University of North Florida, Jacksonville, FL 32224, USA}

\date{\today}

\begin{abstract}
{In this study, we show the divacant substitution of Fe impurities atom produces the formation of an electron spin channel along the arm-chair direction of graphene. This spin channel creates a directional dependent spin exchange between impurities. }Using density functional theory, we simulated the electronic and magnetic properties for a supercell of graphene with spatial variation of the Fe atoms along either the armchair or zig-zag directions. Overall, we find that the exchange interaction between the two Fe atoms fluctuates from ferromagnetic to antiferromagnetic as a function of the spatial distance in the armchair direction. Given the induced magnetic moment and increased density of states at the Fermi level by the surrounding carbon atoms, we conclude that an RKKY-like interaction may characterize the exchange interactions between the Fe atoms. Furthermore, we examined the same interactions for Fe atoms along the zig-zag direction in graphene and found no evidence for an RKKY interaction as this system shows a standard exchange between the transition-metal impurities. Therefore, we determine that the spin channel produced through Fe-substitution in graphene induces a directional-dependent spin interaction, which may provide stability to spintronic and multifunctional devices and applications for graphene.
\end{abstract}

\maketitle


\section{Introduction}

Even after over a decade of high-profile research, graphene continues to provide fascinating results and phenomena\cite{Sier:10, Wolf:14, Novo:12} ranging from semiconductor physics and magnetism to superconductivity and magic angles\cite{Brun:07,Novo:12,cao:18}. The interest in graphene and other two-dimensional (2D) materials is due to their possible technological applications stemming from their electronic properties and high tensile strength, which makes 2D materials excellent candidates for many applications, such as flexible supercapacitors or as a highly sensitive gas detectors\cite{Brun:07,Han:18,Toda:15, Novo:12,cao:18}.

\begin{figure}
  \includegraphics[width=3.25in]{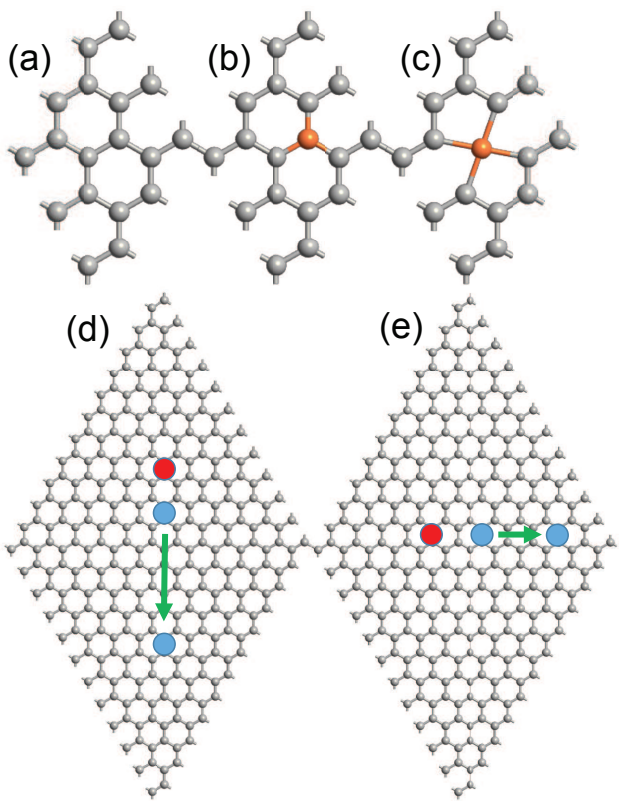}
  \caption{Comparison of the (a) normal graphene structure and the (b) single vacant and (c) di-vacant Fe substitutions. Below shows the range of substituted Fe dimers along the (d) armchair and (e) zigzag directions.}
  \label{struc}
\end{figure}

Graphene has a stable and robust honeycomb lattice (shown in Fig. \ref{struc}(a))\cite{Sier:10, Wolf:14}, where its unique electronic structure and tensile strength strongly suggests electronic applications in memory and logic devices and processes\cite{Baro:06, Fern:07, Drago:17, Kras:11}. These properties have enabled the use of graphene nanoribbons (GNRs) in field-effect transistors, which provide viable technological applications\cite{Baro:06,Fern:07}. Through the utilization of GNRs in quantum dot technology, a large band gap can be introduced producing the possibility of a logic switch\cite{Gucl:14, Jiang:15}.

While nanoscale logic devices are a practical use of graphene, they only take advantage of the charge degree of freedom (DoF). However, in the last decade, there has been a significant push for 2D materials in the realm of spintronics\cite{Han:16,Feng:17}, where the ability for a spintronic device to take advantage of the coupling between magnetic and electronic DoF could develop into new avenues for technological applications\cite{Han:16, Feng:17}. Such a device could be utilized to create great interest in the technology sector due to graphene's inexpensive and practical fabrication.

{Graphene has been shown to gain magnetic moment through introduction of vacancies, adsorption of adtoms, or substitutional impurities, which gives rise to the prospect of spin-based memory storage devices\cite{Leht:04,Yazy:07,kras:09,jaya:10,santo:10,Nair:12,Cerv:09,sant:10,Sarg:11,Mao:08,Hu:08,Cao:10,Sev:08}.  The presence of a magnetic ions in graphene can produce a magnetic coupling between two magnetic moments. Studies examining this effect on graphene have shown to produce a Ruderman-Kittel-Kasuya-Yosida (RKKY) interaction between two magnetic impurities introduced into a single-atomic-layer \cite{santo:10,Crook:15,Houc:17,Koga:11,luan:13}, where an RKKY interaction occurs when two magnetic atoms coupling in a non-magnetic material through the conduction electrons\cite{rude:54,kasu:56,yosi:57}. This effect has been shown via density functional theory (DFT) calculations that the magnetic spin of the metal impurities in graphene changed inversely proportional to the cube of the separation between the two metal impurities \cite{Koga:11}. Furthermore, Some studies have examined the effects of the molecular adsorption on graphene with Fe dimers\cite{tang:17}.}

Recently, it was shown that direct substitution of two transition-metal impurities in place of individual carbon atoms in graphene (shown in Fig. \ref{struc}(b)) resulted in an RKKY interaction\cite{Crook:15} and further showed that these interactions could be used to display the potential of a graphene spintronic device through a change the magnetic spin of the substituted impurities, in a graphene nanoribbon, by a simulated bias voltage \cite{Houc:17}. {Additionally, a previous study done by Seixas et. al. have also shown that there is some magnetic interaction between doped impurities in a similar 2D material phosphorene\cite{Seix:15}}. However, recent studies have shown that a divacant substitution (Fig. \ref{struc}(c)) of metal impurities is the energy-favorable structure over the single vacant substitution\cite{Kras:11,he:14}.

Here, we examine the spatial dependence of electronic and magnetic properties of two magnetic Fe atoms in graphene. Using DFT, we present the electronic band structure, density of states, and magnetization density for Fe atoms separated along both the armchair and zig-zag directions. {We find that the presence of a divacantly-substituted Fe atoms into graphene produces an electron spin channel along the arm-chair direction and mediates a directional-dependent spin exchange between the coupled Fe atoms, where the two Fe atoms seem to couple through an RKKY-like interaction along the armchair direction and a standard exchange in the zig-zag direction. If this effect can be confirmed by experiment, the prospect of spintronic graphene becomes increasing possible.}

\begin{figure}
  \includegraphics[width=3.25in]{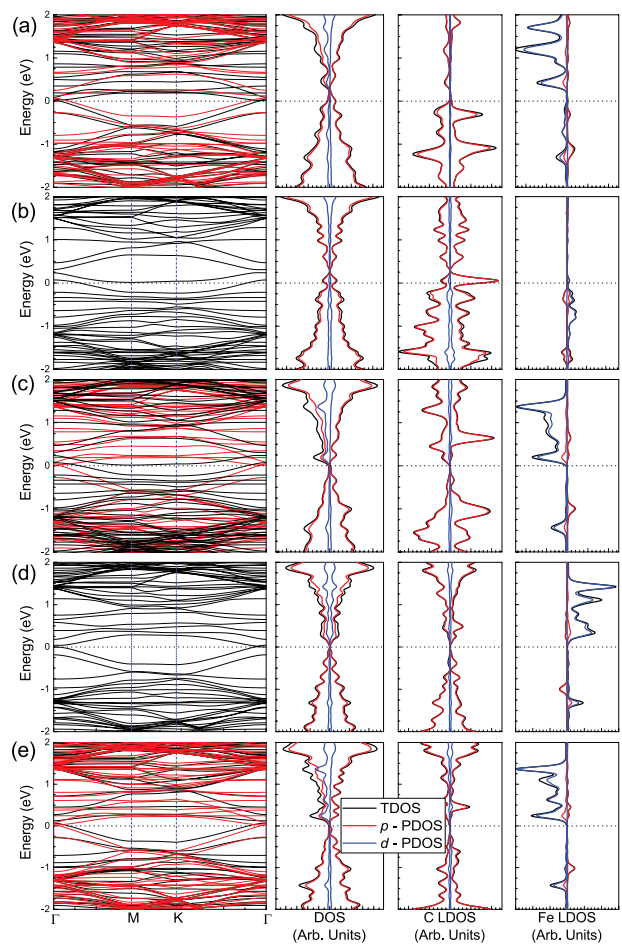}
  \caption{The electronic band structure and density of states for the dimer interactions along the armchair direction with (a) 2 (FM), (b) 4 (AFM), (c) 6 (FM), (d) 8 (AFM), and (e) 10 (FM) carbon separations and exchanges.}
  \label{bs-arm}
\end{figure}

\begin{figure}
  \includegraphics[width=3.25in]{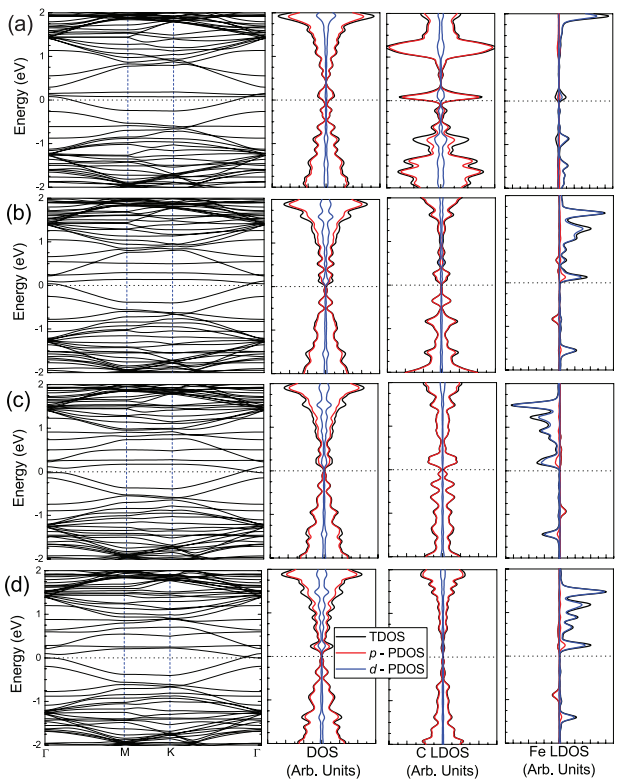}
  \caption{The electronic band structure and density of states for the dimer interactions along the zig-zag direction with (a) 3 (AFM), (b) 5 (AFM), (c) 7 (AFM), and (d) 9 (AFM) carbon separations and exchanges.}
  \label{bs-zz}
\end{figure}

\section{Computational Methodology}

Calculations were performed using DFT through the utilization of QuantumATK\cite{quantumwise,dft,QATK}. Starting with a 12x12 supercell of graphene (216 atoms), two C atoms were replaced by a single Fe atom. Another Fe atom was then inserted into graphene in the same manner at varying distances from the original Fe atom. This spatial distance was performed along both the zig-zag and armchair directions, shown in Figure \ref{struc}(d) and (e), respectively. Using a geometry optimization, the supercell was relaxed to find the ground state configuration. {The results were then compared to previous literature to assure consistency\cite{sant:10,santo:10}. Our choice of Fe was due to the experimental work on the energy favorable configurations for Fe dimers shown Ref. [\onlinecite{he:14}]. Stability is a critical component since studies on 5d transition metals, adtoms, and vacancies have shown to be unstable and produce ion migration using tungsten\cite{meyer:08,cret:10,bouk:09}.}

{Using a Spin-polarized Generalized Gradient Approximation (SGGA) in the {Perdew, Burke, and Ernzerhof (PBE) functional\cite{sole:02,perd:96}}, which utilizes a norm-conserving pseudopotential of Fe including nonlinear core corrections. Our study was applied using a 10x10x1 k-point sampling, and a Hubbard potential of 4 eV was placed on the Fe 3$d$-orbitals produce localized magnetic moments.} The electronic band structure, density of states (DOS), and magnetization density were determined. Through an examination of the total energy for the ferromagnetic (FM) and antiferromagnetic (AFM) configurations, the ground state was established. With the ground state known, we then analyzed the electronic structure to determine the nature of the exchange coupling.

\section{Results and Discussion}

\subsection{Electronic Band Structure}

{Figure \ref{bs-arm} shows the electronic band structure and DOS for the divacant substitution of graphene with Fe dimers along the armchair direction for impurity separations of 2, 4, 6, 8, and 10 carbon atoms. The band structure contains both spin up (black) and spin down (red) bands when the ground state has a FM polarization, but the bands are degenerate in the AFM polarization.} Furthermore, the DOS for each configuration is paired alongside the band structure, where we present the total and partial DOS, as well as the local DOS (LDOS) for C, and the LDOS of Fe (moving left to right). 

{Figure \ref{bs-zz} shows the electronic band structure and density of states (DOS) for the divacant substitution of graphene with Fe dimers along the zig-zag direction for impurity separations of 3, 5, 7, and 9 carbon atoms. Here, all ground states are AFM. Therefore, the spin up and down bands are degenerate.} The DOS for each configuration is paired alongside the band structure, where we present the total and partial DOS, as well as the local DOS (LDOS) for C, and the LDOS of Fe (moving left to right). 

{The electronic band structure for the arm-chair and zig-zag configurations show a clear difference in the magnetic interactions within the system. Both seems to have the presence of conduction electrons. However, only the arm-chair direction presences the alternating magnetic interactions typically observed for RKKY interactions. On contrary, the zig-zag direction does not change in magnetic character, which is {indicative} of a standard exchange model. Therefore, the simple presence of conduction electrons does not mean that a RKKY interaction will be present. To understand this further, we will examine the magnetic interactions and the electron density.}

\begin{figure}
  \includegraphics[width=3.5in]{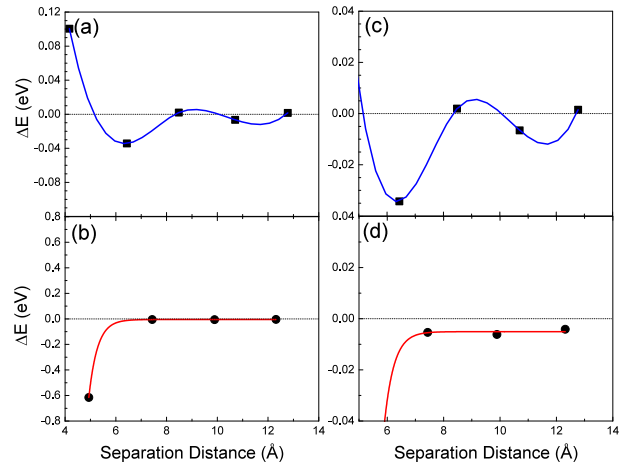}
  \caption{The change in energy between ferromagnetic (positive) and antiferromagnetic (negative) configurations for the (a) armchair and (b) zigzag directions as a function of separation distance. The change in energy is directly proportional to the exchange interaction. Here, the armchair configuration seems to be dominated by a RKKY-like interaction, which is characterized by the oscillatory nature of the spin exchange, while the zigzag configuration seems to only be dominated by standard exchange.{As the separation between the impurities is increased, the exchange energy is reduced (shown in (c) and (d)).}}
  \label{DeltaE}
\end{figure}

\begin{figure*}
  \includegraphics[width=6.5in]{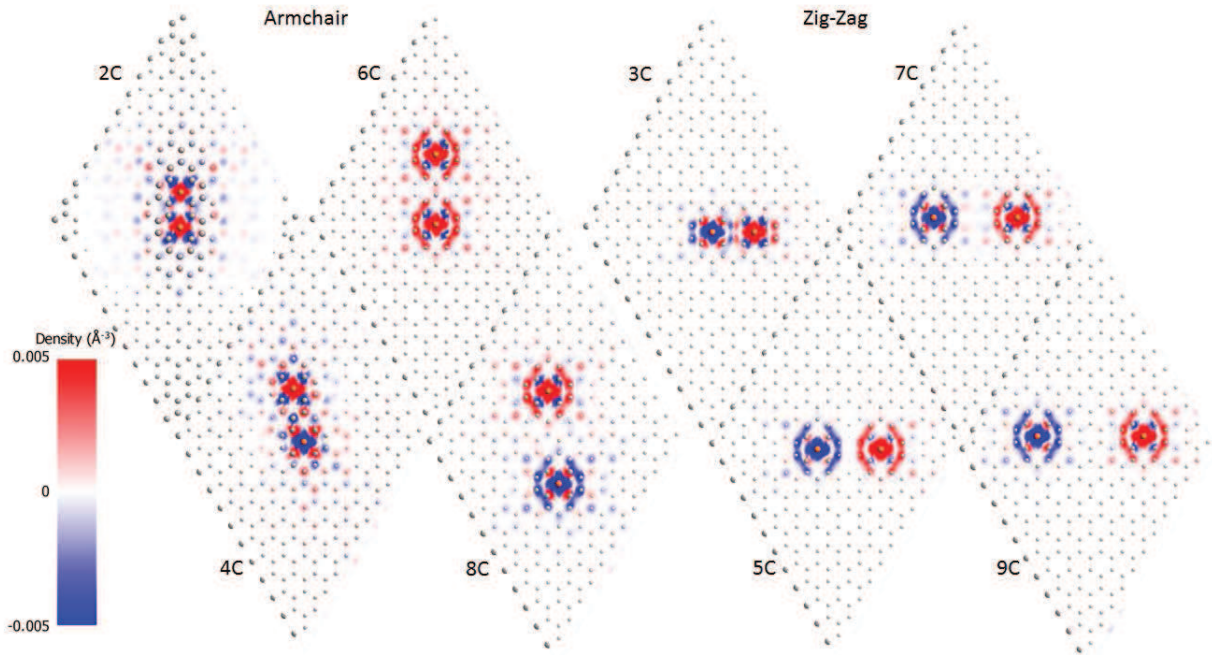}
  \caption{The magnetization density between the spin up and down channels for the Fe dimer interactions along the armchair and zigzag directions. At large separation distances, the armchair configuration seems to transition from a RKKY dominated interaction to that of a standard exchange. This is characterized by a shift in the orbital interactions.}
  \label{mag}
\end{figure*}

\subsection{Magnetic Interactions}

To understand the magnetic and electronic properties of Fe atoms divacantly substituted into a supercell of graphene, the exchange characteristics of the magnetic dimer in graphene can be considered by a standard exchange spin Hamiltonian,
\begin{equation}
H = -J \vec{S_1}\cdot \vec{S_2},
\end{equation}
where J is the exchange interaction between the two spins. In a standard exchange quantum spin dimer, the change in energy between the two states J\cite{houc:15}, where you get a FM ground state for J $>$ 0 and an AFM ground state for J $<$ 0. Therefore, through an examination of the energy difference of the FM and AFM configurations, one can determine the ground state configuration and estimate the exchange interaction.

Figure \ref{DeltaE} shows the change in energy between the FM and AFM configurations. For the armchair configuration (Fig. \ref{DeltaE}(a) and (c)), the change in energy shows a distinct oscillation between FM and AFM ground states, where the zig-zag configuration (Fig. \ref{DeltaE}(b) and (d)) is consistently the below zero and indicates constant AFM exchange. This difference in exchange shows a directional-dependent exchange in graphene, where the armchair configuration is consistent with that of an RKKY interaction, which has been shown in the single vacant case as well. However, the zig-zag configuration only produces a standard exchange between the impurity atoms.

An analysis of the electronic structure and local density of states for the carbon atom indicates a slight metallicity between the two Fe impurities. Therefore, given alternating magnetic ground state shown in Fig. \ref{DeltaE}(a) and the modest density of states in the linking carbon atoms, it is likely that an RKKY interaction exists between Fe impurities along the armchair direction, which is in agreement with previous calculations on metal impurities substituted in graphene\cite{Koga:11}.

{\subsection{Electron spin channel}}

{Figure \ref{mag} shows the magnetization density for Fe atoms substituted along the armchair and zig-zag directions with increasing spatial distance. An analysis of the density plot shows the magnetic spin of the Fe atoms and their influence on the surrounding carbon atoms as the two Fe atoms are separated along both directions. The illustration of the magnetization density indicates the presence of a spin channel that form along the arm-chair direction of graphene.}

{Along the armchair direction, we find that the orbital interactions between the magnetic impurities changes and alternates between FM and AFM. Furthermore, the electron distribution in the 2 and 4 carbon separation cases also produce a different pattern than the others. {This is likely due to a small, but noticeable, distortion in the lattice, which is illustrated by the faded regions in Fig. \ref{mag}.} However, there appears to be a clear competition between RKKY-like interactions through the conduction electrons and a standard Heisenberg interaction through orbital interactions.}

{Examining magnetization density along the zig-zag direction, the configuration of the two Fe impurities does not exhibit the same characteristics shown by the arm-chair direction. Here, the magnetic ground state along the zig-zag direction remains AFM, which is consistent with the electronic structure analysis. Therefore, the nature of these interactions are likely a standard orbital exchange as was shown through an examination of the ground state energy between FM and AFM interactions (Fig. \ref{DeltaE}(b)).}

\begin{figure}
  \includegraphics[width=2.75in]{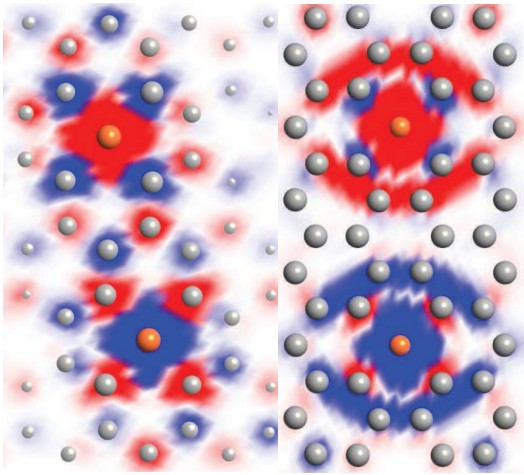}
  \caption{{The development of a spin channel through the conduction electrons in the armchair directions is evident through the antiferromagnetic ordering through graphene. However, in the zig-zag case, the orbitals along the zig-zag direction produce a spin shield that seems to block exchange through the RKKY interaction. Therefore, the zig-zag direction couples as normal orbital exchange.}}
  \label{mag-compare}
\end{figure}

{Within this picture, the nature of the magnetic coupling can be better understood. The divacant substitution of Fe into the lattice produces a break in the local symmetry and forms an electron spin channel around the embedded spin. This is clearly shown when the impurity atoms are far apart. As they are brought closer together, small lattice distortions blur the exchange pathway. As shown in Fig. \ref{mag-compare}, the spin channel is directed along the arm-chair direction, which allows for easier interaction of the spins using the conduction electrons, which enables an RKKY-like interaction. We call it RKKY-like because the system is a combination of RKKY interactions and standard exchange. This also explains why the interactions along the zig-zag direction does not exhibit RKKY-like features. The presence of a spin channel induces a magnetization shield around the impurity atom, which essentially ``blocks" the conduction electron contribution and the system remains AFM.}

\section{Conclusion}

Overall, we show that the presence of divacantly-substituted atoms in graphene produces a directional-dependent magnetic exchange due to the formation of an electron spin channel that favors conduction electrons along the arm-chair direction. Divacantly-substituted Fe atoms were placed into a graphene sheet and along either the armchair or zig-zag directions. Using DFT, the electronic and magnetic properties for the Fe atoms were determined for various spatial separations. An analysis of the exchange energy shows that graphene produces a directional-dependent magnetic exchange. Interactions along the zig-zag direction seem to be governed by a standard exchange, while the interactions along the arm-chair direction produce an RKKY-like exchange.  

We show that systematic placement of magnetic impurities not only induces magnetism into the graphene lattice, but the presence of the spin channel allows for the directional dependence in graphene to form. The interaction through the conduction electrons along the zig-zag are blocked by the magnetization shield of the spin channel. This blockade makes the zig-zag direction rely on standard orbital interactions, rather than conduction electron interactions. Previous studies have only shown that graphene has either an RKKY or standard orbital exchange. However, our study shows that the interactions depends on the orbital manipulations of the lattice.

Overall, this dichotomy of interactions may provide an avenue for which graphene can be used in the development of a spintronic device as it has implications for both technological and fundamental understanding of magnetism in graphene. Further studies hope to show that it may be possible to influence electrons through voltage gates and take advantage of the magnetic spin of the impurities.

\section*{Acknowledgements}

The authors acknowledge useful and fruitful discussions with J.-X. Zhu, T. Ahmed, F. Ronning, and J. Mendoza-Cortes. R.P. and J.T.H recognize support from the Institute for Materials Science at Los Alamos National Laboratory as well as faculty development grant from Academic Affairs at the University of North Florida.


\begin{thebibliography}{GPH}

\bibitem{Sier:10} U. Sierra, P. Alvarez, C. Blanco, M. Granda, R. Santamaria, and R. Menendez, New alternatives to graphite for producing graphene materials, Elsevier: Carbon {\bf 93}, 812 (2010).

\bibitem{Wolf:14} E. Wolf and SpringerLink, Applications of Graphene An Overview, Cham: Springer International Publishing: Imprint: Springer, 19 \& 51 (2014).

\bibitem{Novo:12} K. S. Novosolev, V. I. Fal'ko, L. Colombo, P. R. Gellert, M. G. Schwab and K. Kim, A Roadmap for Graphene, Nature {\bf 490}, 192–200 (2012).

\bibitem{Brun:07} U. Bruno and N. A. H. Castro, Superconducting States of Pure and Doped Graphene, Phys. Rev. Letters {\bf 98}, 146801 (2007).

\bibitem{cao:18} Y. Cao, V. Fatemi, S. Fang, K. Watanabe, T. Taniguchi, E. Kaxiras, and P. Jarillo-Herrero, Unconventional superconductivity in magic-angle graphene superlattices, Nature {\bf 556}, 43 (2018).



\bibitem{Han:18} Y. Han, Y. Ge, Y. Chao, C. Wan, and G. G. Wallace, Recent Progress in 2D Materials for Flexible Supercapacitors, Journal of Energy Chemistry {\bf 27}, 57 (2018).

\bibitem{Toda:15} K. Toda,  R. Furue, and  S. Hayami, Recent progress in applications of graphene oxide for gas sensing: A review, Elsevier: Analytica Chimica Acta {\bf 878}, 43 (2015)

\bibitem{Baro:06} V. Barone, O. Hod and G. Scuseria, Electronic structure and stability of semiconducting graphene nanoribbons, Nano Letters {\bf 6(12)}, 2748 (2006).

\bibitem{Fern:07} J. Fernández-Rossier, J. J. Palacios and L. Brey, Electronic structure of gated graphene and graphene ribbons, Phys. Rev. B {\bf 75(20)} (2007).  

\bibitem{Drago:17} M. Dragoman, D. Dragoman and SpringerLink, 2D Nanoelectronics Physics and Devices of Atomically Thin Materials, Cham: Springer International Publishing: Imprint: Springer ebook, 161 (2017).


\bibitem{Kras:11} A. V. Krasheninnikov and R. M. Nieminen, Attractive Interaction Between Transition-Metal Atom Impurities and Vacancies in Graphene: A First-Principles Study, Theoretical Chemistry Accounts {\bf 129}, 625 (2011).

\bibitem{he:14} Z. He, K. He, A.W. Robertson, A.I. Kirkland, D. Kim, J. Ihm, E. Yoon, G.-D. Lee, and J.H. Warner, Atomic Structure and Dynamics of Metal Dopant Pairs in Graphene. Nano Lett. {\bf 14}, 3766 (2014).

\bibitem{Gucl:14} A. D. G\"{u}cl\"{u}, P. Potasz, M. Korkusinski, P. Hawrylak and Springerlink, Graphene Quantum Dots, Berlin, Heidelburg: Springer Berlin Heidelberg: Imprint: Springer, 29 (2014).

\bibitem{Jiang:15} L. Jiang, X. Zhao, and Y. Zheng, RKKY interaction in graphene with a line defect, Journal of Physics: Condensed Matter {\bf 27}, 10 (2015).

\bibitem{Han:16} Han, W., Perspectives for Spintronics in 2D Materials, APL Materials {\bf 4} (2016).


\bibitem{Feng:17} Y. Feng, L. Shen, M. Yang, A. Wang, M. Zeng, Q. Wu, S. Chintalapti and C. R. Chang, Prospects of spintronics based on 2D materials, Wiley Interdisciplinary Reviews: Computational Molecular Science {\bf 7(5)} (2017).

\bibitem{Leht:04} P.P Lehtinen, A. S. Foster, Y. Ma, A. V. Krasheninnikov, and R. M. Nieminen, Irradiation-Induced Magnetism in Grphite: A Density Functional Study, Physical Review Letters {\bf 93}, 187202 (2004)

\bibitem{Yazy:07} O. V. Yazyev and L. Helm, Defect-induced magnetism in graphene, Phys. Rev. B {\bf 75}, 125408 (2007).

\bibitem{kras:09} {A. V. Krasheninnikov, P. O. Lehtinen, A. S. Foster, P. Pyykko, and R. M. Nieminen, Embedding Transition-Metal Atoms in Graphene: Structure, Bonding, and Magnetism, Phys. Rev. Lett. {\bf 102}, 126807 (2009).}

\bibitem{jaya:10} {T. Jayasekera, B. D. Kong, K. W. Kim, and M. Buongiorno Nardelli, Band Engineering and Magnetic Doping of Epitaxial Graphene on SiC (0001), Phys. Rev. Lett. {\bf 104}, 146801 (2010).} 



\bibitem{Nair:12} R. R. Nair, M. Sepioni, I. L. Tsai, O. Lehtinnen, J. Keinonen, A. V. Krasheninnikov, T. Thomson, A. K. Geim and I. V. Grigorieva, Spin-half paramagnetism in graphene induced by point defects, Nature Physics {\bf 8}, 199 (2012).

\bibitem{Cerv:09} J. Cervenka, M. I. Katsnelson, and C.F.J. Flipse, Room-temperature ferromagnetism in graphite driven by two-dimensional networks of point defects, Nature Physics {\bf 5}, 840 (2009).

\bibitem{sant:10} {E. J. G. Santos, A. Ayuela and D. Sánchez-Portal, First-principles Study of Substitutional Metal Impurities in Graphene: Structural, Electronic, and Magnetic Properties, New Journal of Physics {\bf 12}, 053012 (2010).}



\bibitem{Sarg:11} {M. Sargolzaei and F. Gudarzi, Magnetic Properties of Single 3d Transition Metals Absorbed on Graphene and Benzene: A Density Functional Theory Study, Journal of Applied Physics {\bf 110}, 064303 (2011).}

\bibitem{Mao:08} {Y. Mao, J. Yuan and J.X. Zhong, Density Functional Calculation of Transition Metal Adatom Adsorption on Graphene, Journal of Physics: Condensed Matter {\bf 20}, 115209 (2008).}


\bibitem{Hu:08} {L. Hu, X. Hu, X. Wu, C. Du, Y. Dai and J. Deng, Density Functional Calculation of Transition Metal Adatom Adsorption on Graphene, Physica B: Condensed Matter {\bf 405}, 3337 (2010).}

\bibitem{Cao:10} {C. Cao, M. Wu, J. Z. Jiang and H. P. Cheng, Transition Metal Adatom and Dimer Adsorbed on Graphene: Induced Magnetization and Electronic Structures, Phys. Rev. B {\bf 81}, 205424 (2010).}

\bibitem{Sev:08} {H. Sevincli, M. Topsakal, E. Durgun and S. Ciraci, Electronic and Magnetic Properties of 3d Transition-Metal Adatom Adsorbed Graphene and Graphene Nanoribbons, Phys. Rev. {\bf 77}, 195434 (2007).}



\bibitem{santo:10} {E. J. G. Santos, D. S\'anchez-Portal, and A. Ayuela, Magnetism of substitutional Co impurities in graphene: Realization of single $\pi$ vacancies, Phys. Rev. B {\bf 81}, 125433 (2010).}

\bibitem{luan:13} {H.-X. Luan, C.-W. Zhang, S.-S. Li, R.-W. Zhanga, and  P.-J. Wanga, First-principles study on ferromagnetism in W-doped graphene, RSC Adv. {\bf 3}, 26261 (2013)} 


\bibitem{Crook:15} C. B. Crook, C. Constantine, T. Ahmed, J.X. Zhu, A. V. Balatsky and J. T. Haraldsen, Proximity-induced Magnetism in Transition-metal Substituted Graphene, Scientific Reports {\bf 5}, 12322 (2015).


\bibitem{Koga:11} E. Kogan, RKKY interaction in graphene, Phys. Rev. B {\bf 84}, 115119 (2011).

\bibitem{Houc:17} G. Houchins, C. B. Crook, J. X. Zhu, A. V. Balatsky and J. T. Haraldsen, Voltage-dependent Spin Flip in Magnetically Substituted Graphene Nanoribbons: Towards the Realization of Graphene-based Spintronic Devices, Phys. Rev. B {\bf 95}, 155450 (2017).

\bibitem{rude:54} M. A. Ruderman and C. Kittel, Indirect Exchange Coupling of Nuclear Magnetic Moments by Conduction Electrons, Phys. Rev. {\bf 96}, 99 (1954). 

\bibitem{kasu:56} T. Kasuya, A Theory of Metallic Ferro- and Antiferromagnetism on Zener's Model, Prog. Theor. Phys. {\bf 16}, 45 (1956). 



\bibitem{yosi:57} K. Yosida, Magnetic Properties of Cu-Mn Alloys, Phys. Rev. {\bf 106}, 893 (1957).

\bibitem{tang:17} {Y. Tang, H. Chai, W. Chen, X. Cui, Y. Ma, M. Zhaob,  and  X. Dai,Theoretical study on geometric, electronic and catalytic performances of Fe dopant pairs in graphene, Phys. Chem. Chem. Phys. {\bf 19}, 26369 (2017).} 

\bibitem{Seix:15} L. Seixas, A. Carvalho, and H. Castro Neto, Atomically thin dilute magnetism in Co-doped phosphorene, Phy. Rev. B {\bf 91}, 155138 (2015)

\bibitem{Gan:08} Y. Gan, L. Sun and F. Banhart, One and Two Dimensional Diffusion of Metal Atoms in Graphene, Small {\bf 4(5)}, 587 (2008).

\bibitem{quantumwise} QuantumATK version 2017.2, Synopsys QuantumATK (https://www.synopsys.com/silicon/quantumatk.html)


\bibitem{dft} S. Smidstrup, D. Stradi, J. Wellendorff, P. A. Khomyakov, U. G. Vej-Hansen, M-E. Lee, T. Ghosh, E. Jónsson, H. Jónsson, and K. Stokbro, First-principles Green's-function method for surface calculations: A pseudopotential localized basis set approach, Phys. Rev. B {\bf 96}, 195309 (2017).

\bibitem{QATK} {S. Smidstrup1, T. Markussen1, P. Vancraeyveld1, J. Wellendorff1, J. Schneider1, T. Gunst1, B. Verstichel1, D. Stradi1, P. A. Khomyakov1, U. G. Vej-Hansen1, QuantumATK: An integrated platform of electronic and atomic-scale modelling tools, J. Phys.: Condens. Matter in press (2019).}

\bibitem{sole:02} J. M. Soler, E. Artacho, J. D. Gale, A. García, J. Junquera, P. Ordejón, and D. Sánchez-Portal, J. Phys. Condens. Matter {\bf 14}, 2745 (2002).

\bibitem{meyer:08} {J. C. Meyer, C. Kisielowski, R. Erni, M. D. Rossell, M. F. Crommie, and A. Zettl, Direct Imaging of Lattice Atoms and Topological Defects in Graphene Membranes, Nano Lett. {\bf 8} 3582 (2008).}

\bibitem{cret:10} {O. Cretu, A. V. Krasheninnikov, J. A. Rodr\'iguez-Manzo, L. Sun, R. M. Nieminen, and F. Banhart, Migration and Localization of Metal Atoms on Strained Graphene, Phys. Rev. Lett. {\bf 105}, 196102 (2010).}

\bibitem{bouk:09} {D. W. Boukhvalova and M. I. Katsnelson, Destruction of graphene by metal adatoms, Appl. Phys. Lett. {\bf 95}, 023109 (2009).}

\bibitem{perd:96} J.P. Perdew, K. Burke, and M. Ernzerhof, Generalized Gradient Approximation Made Simple, Phys. Rev. Lett. {\bf 77},  3865-3868 (1996).


\bibitem{houc:15} {G. Houchins and J. T. Haraldsen, Generalization of polarized spin excitations for asymmetric dimeric systems, Phys. Rev. B {\bf 91}, 014422 (2015).}




\end{thebibliography}
\end{document}